# Electrical Circuit Modelling of Nanofluidic Systems

John Sebastian and Yoav Green*

Department of Mechanical Engineering, Ben-Gurion University of the Negev, Beer-Sheva 8410501, Israel

JS ORCID number: 0000-0003-3430-1349
YG ORCID number: 0000-0002-0809-6575
* yoavgreen@bgu.ac.il

**Abstract**

Nanofluidic systems exhibit transport characteristics that have made technological marvels such as desalination and energy harvesting possible by virtue of their ability to influence small currents due to selective ion transport. Traditionally, these applications have relied on nanoporous membranes whose complicated geometry impedes a comprehensive understanding of the underlying physics. To bypass the associated difficulties, we consider the simpler nanochannel array and elucidate the effects of interchannel interactions on the Ohmic response. We demonstrate that a nanochannel array is equivalent to an array of mutually independent but identical unit-cells whereby the array can be represented by an equivalent electrical circuit of resistances connected in a parallel configuration. Our model is validated using numerical simulations and experiments. Our approach to modeling nanofluidic systems by their equivalent electrical circuit provides an invaluable tool for analyzing and interpreting experimental measurements.

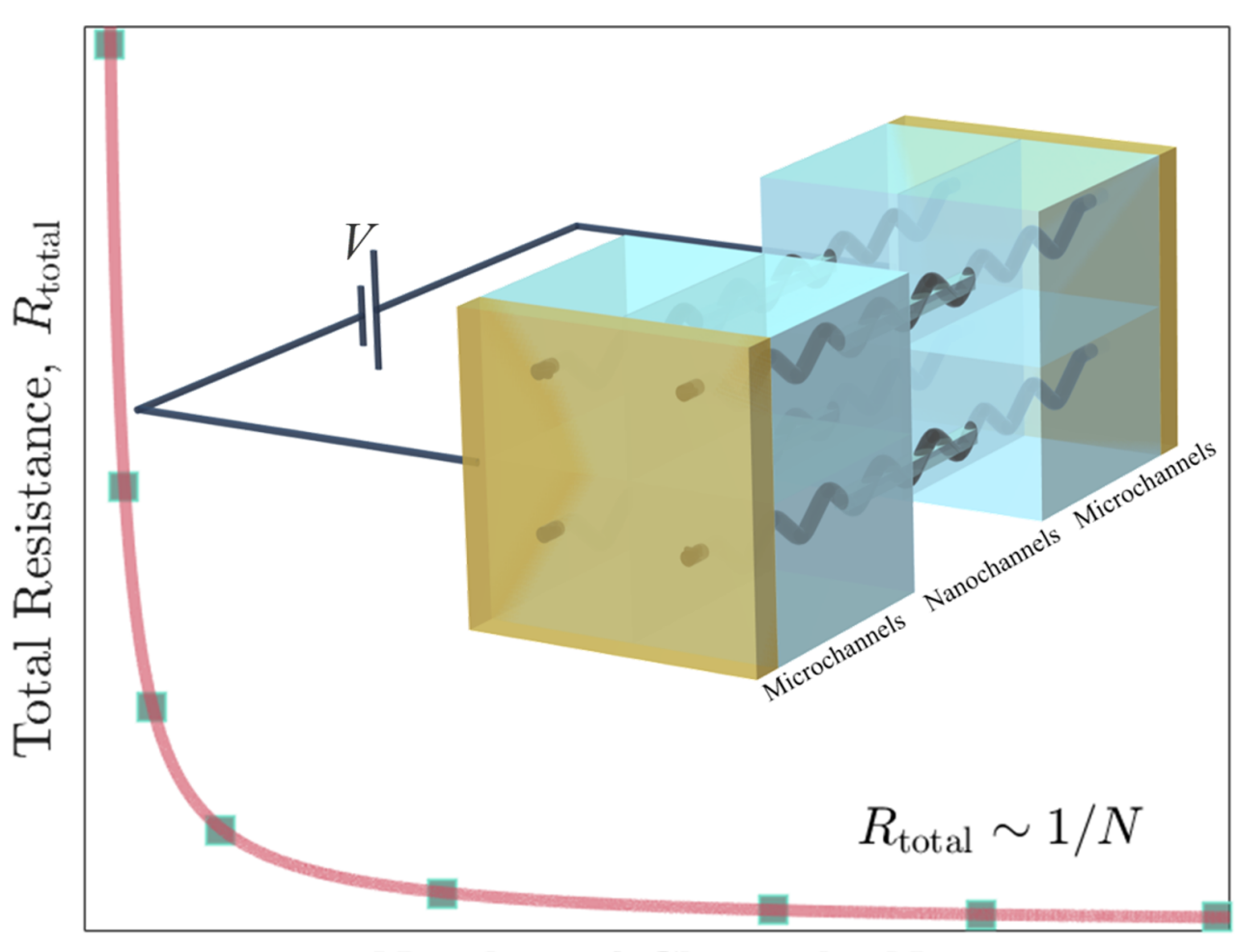

## 1. Introduction

With the ever-increasing demand for fresh-water and clean renewable energy[1] combined with the ever-increasing appearances of global warming effects, it has never been more apparent and more urgent to improve the performance (efficiency, power input, etc.) of water desalination and energy harvesting systems that utilize nanoporous ion-selective membranes. In recent decades, research has been divided into two main thrusts: one, improving the material properties of membranes, and two, improving our fundamental understanding of the phenomena occurring at these very small scales. To achieve the first objective, scientists tune the material properties of macroscopically large membranes and compare how they impact their electric response and process efficiencies. Towards the second objective, the much simpler, if not entirely realistic, single nanochannel setup is adopted. This scenario allows the probing of the fundamental physics of various nanofluidic and electrokinetic effects to a higher resolution. However, several phenomena that emerge from the scaling up from a microscopically small single nanochannel system to a macroscopically large membrane system are yet to be fully expounded. Here we will demonstrate how the nanochannel array serves as the intermediate of these two different scenarios and how the response varies as the number of channels increases from one (single-channel system) to an arbitrary number, $N$.

At sufficiently low concentrations and large surface charge densities, the electric double layers within the pores overlap, such that the membranes are ion-selective and capable of perfect coion exclusion (discussed further below). Once perfect coion exclusion has been achieved, electrodialysis[2,3] (ED)

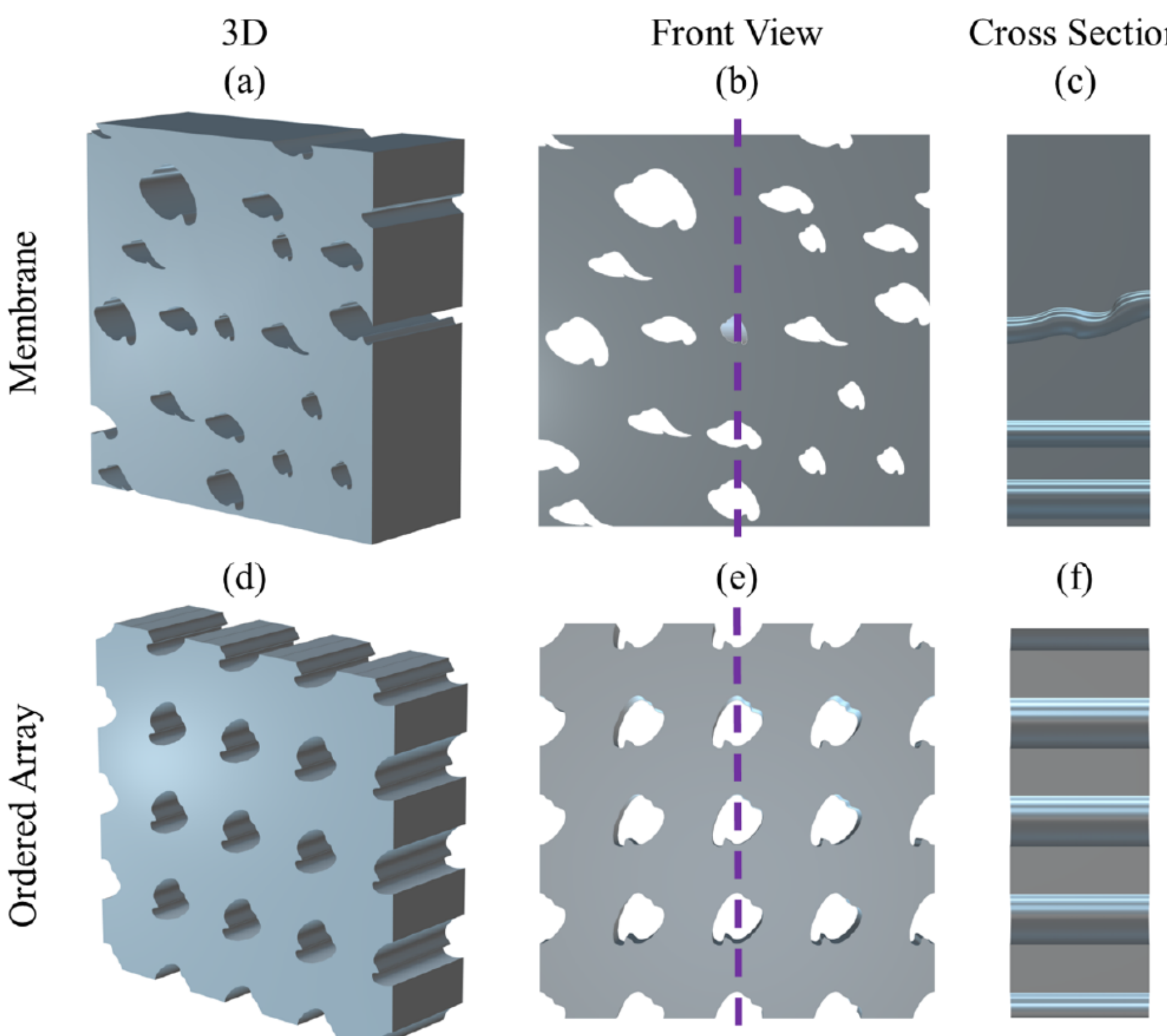

**Figure 1** (a) Three-dimensional view, (b) front-view, and (c) the cross-section of the nanoporous membrane along the purple line in (b). (d) Three-dimensional view, (e) front-view, and (f) the cross-section of an *ordered* array of nanopores along the purple in (e). The geometry of the ordered nanopore array is free of the irregular porosity and random tortuosity characteristic of conventional nanoporous membranes.

and reverse electrodialysis[3–6] (RED) are robust processes that are virtually independent of the material itself. The feasibility of ED and RED employing nanoporous membranes results from their large pore densities that allow for relatively large fluxes – this is essentially a parallelization process whereby all the pores participate in ion transport. However, thus far, improving the efficiency of such systems has primarily relied on trial-and-error and empirical investigations of material properties and geometric configurations. The traditional reliance on this approach stems from the difficulties arising from the irregular porosity and random tortuosity of the membrane [see **Figure 1**(a)-(c) for membrane schematic] – these geometric features do not allow for a straightforward analysis of the fundamental and microscopic physics occurring at the smallest scales. Consequently, most analyses have been limited to simplified one-dimensional (1D) models that do not account for crucial microscopic details.

In the foregoing decades, the *single* nanochannel system[7–10] was introduced as the simplest tractable model for its larger cantankerous counterpart – the membrane. Similar to the membrane, nanochannels are ion-selective at low concentrations, exhibiting perfect coion exclusion. Importantly, nanochannels benefit from two additional advantages. First, their simple geometry is easy to fabricate, easy to comprehend, and easy to analyze (experimentally and theoretically), whereby the fundamental physics becomes more apparent. Second, additional applications that cannot be realized with conventional membranes are immediately contrived with small nanochannels. The well-defined, deterministic geometries of these engineered nanochannels make them highly amenable to chemo and bio-sensing[11,12], DNA sequencing[13], and fluid-based electrical circuits[14–17]. Nonetheless, while single-channel systems are favorable for some applications, they are inadequate for others. Specifically, the disadvantage of the single nanochannel is that the relatively low fluxes make it impractical to use without scaling up – their low throughput makes them irrelevant for realistic ED and RED applications.

The natural bridge between macroscopically large membrane systems and microscopically small single nanochannel systems is the ordered nanochannel array system[18–21] [**Figure 1**(d)-(f)]. The nanochannel array promises the advantages of both systems: regularized and simple geometry combined

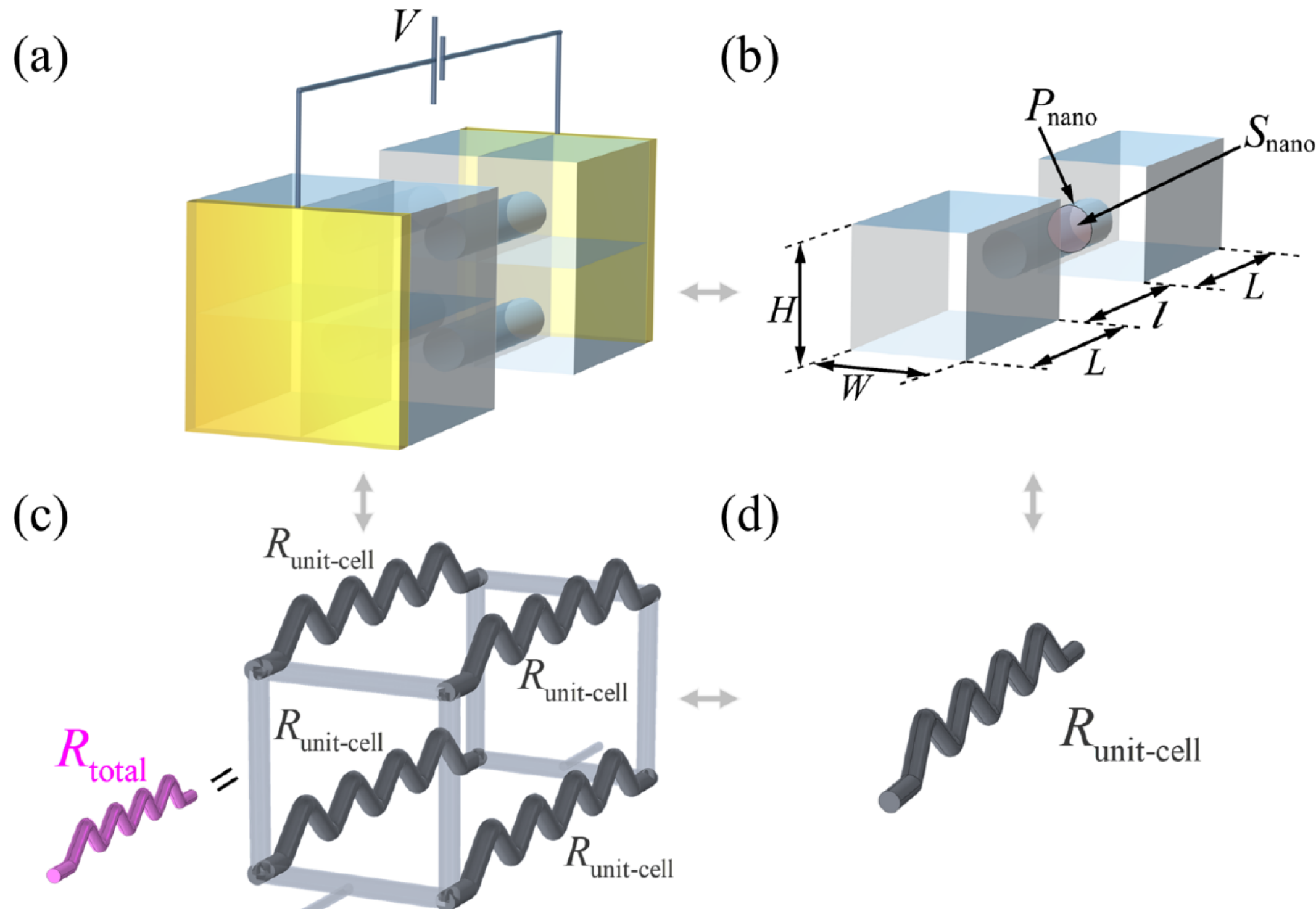

**Figure 2** (a) Illustration of a 2 × 2 array of a microchannel-nanochannel system. The domain shown is formed by the inlet and outlet reservoirs bridged by four nanochannels filled with an electrolyte of concentration, $c_0$. (b) Representative unit-cell comprised of two microchannels connected by a single nanochannel with perimeter, $P_{nano}$, and cross-section area, $S_{nano}$. (c) The equivalent circuit of the 2 × 2 array in (a) is a circuit of 4 unit-cell resistances (b) connected in parallel. This circuit can further be abstracted as the total resistance $R_{total}$ [Eq. (1)] of the array. (d) The unit-cell resistance, $R_{unit\text{-}cell}$ [Eq. (2)], is discussed in the main text.

with full potential for parallelization to upscale the fluxes. However, the physics of nanochannel arrays is yet to be discerned; namely, the interactions between multiple nanochannels are not fully understood. The purpose of this paper is to elucidate the governing physics and delineate the electrical response of multichannel systems by demonstrating that these complicated systems can be represented by a simple, equivalent electrical circuit. Specifically, we will show that an array of microchannel-nanochannels [**Figure 2**(a)] is comprised of independent "unit-cells" [**Figure 2**(b)], such that the total resistance of the system is equivalent to an electrical circuit of resistances connected in parallel configuration [**Figure 2**(c)] composed of "unit-cell" resistors, [**Figure 2**(d), $R_{unit\text{-}cell}$ is discussed thoroughly further below].

Our goal is to show that an ordered nanochannel array leads to the partitioning of the fluidic domain composed of the reservoirs and individual nanochannels into an ordered array of unit-cells. To that end, this paper is divided as follows. In Sec. 2, we discuss the concept of the unit cell and how to utilize it when parallelizing the electrical circuit of nanochannel arrays. In Sec. 3, we present numerical simulations and experimental results that confirm our theoretical prediction. Sec. 4 reviews and compares our model to other suggested models, as well as discusses the outcomes of our results. We conclude with short remarks in Sec. 5.

## 2. Electrical resistance of nanochannels and microchannels

Section 2.1 presents numerical simulations proving that adjacent unit-cells do not exchange flux. Section 2.2 discusses the outcome of this result – namely, why the parallel circuit of **Figure 2**(c) is justifiable. Section 2.3 discusses the unit-cell resistances shown in **Figure 2**(d).

### 2.1. Numerical simulations

To elucidate the electrical response of nanochannel arrays, we simulated both 2D and 3D arrays of nanochannels connecting two adjacent microchannels (**Figure 3**). The geometry of all the nanochannels and all the microchannels are kept constant such that the array is simply a multiplication of the "unit-cell".

First, we leverage our understanding from single channel systems. Namely, in single channel systems, it is clear that the walls provide boundary conditions of no-flux for electric field lines, ionic flux, electrical current density, etc. Mathematically, such a boundary condition can be written as $\boldsymbol{j} \cdot \boldsymbol{n} = 0$ where $\boldsymbol{j}$ is the generalized flux and $\boldsymbol{n}$ is the unit vector normal to the

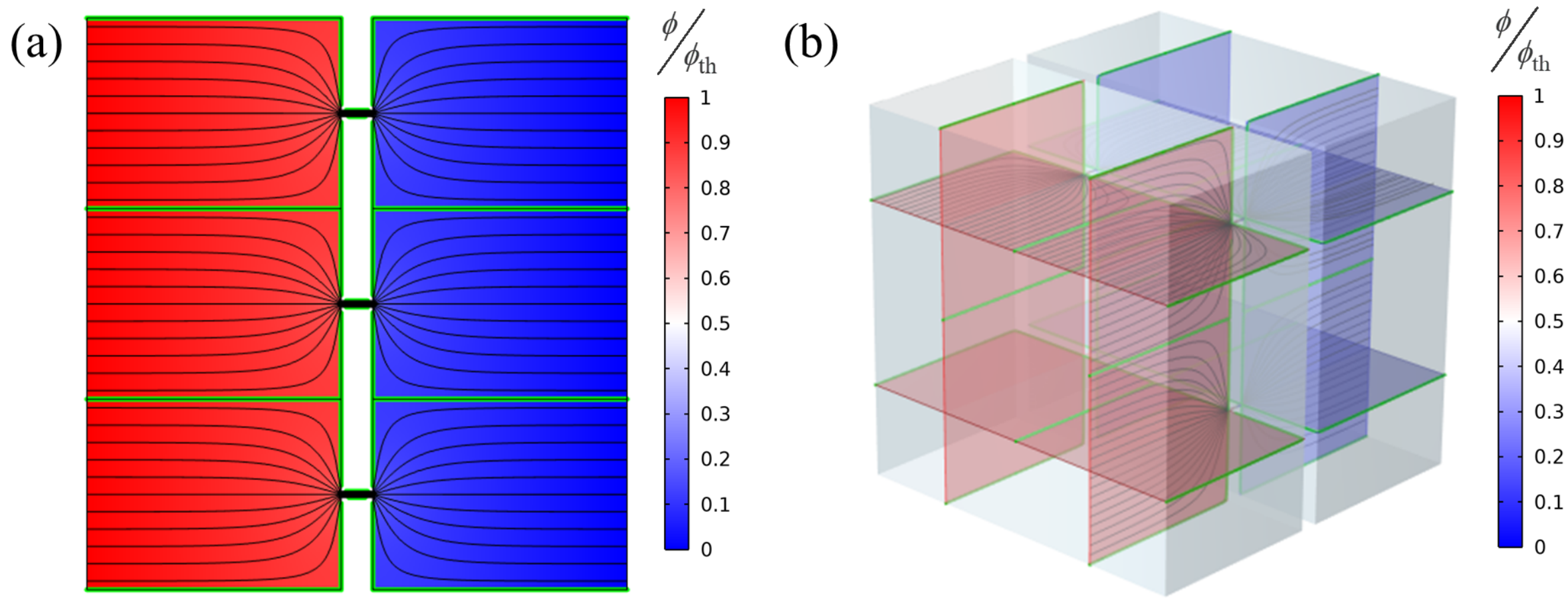

**Figure 3** (a) The 2D electric potential distribution, $\phi$, from numerical simulations. For demonstration purposes, we plot only three unit-cells out of the $N = 50$ unit-cells calculated. (b) The 3D electric potential distribution, $\phi$, from numerical simulations of a cylindrical nanopore. In both plots, black curves are the streamlines of the electric field, $\boldsymbol{E} = -\nabla\phi$. The streamlines denoting the formation of unit-cells are highlighted in green. The potentials have been normalized by the thermal potential $\phi_{\text{th}} = \Re T/F$ (at room temperature $T = 298K$). See Supporting Information[22] for geometric details.

boundary. However, the boundary condition $\boldsymbol{j} \cdot \boldsymbol{n} = 0$ is not unique to solid walls. It also describes the boundary condition for planes of symmetry. Thus, for the arrays in **Figure 2**, it is expected that inner lines/planes of symmetry will form to ensure flux conservation within every independent unit cell. If so, unit-cells are, indeed, mutually independent "building blocks" of the array.

In our numerical simulations, we dictate boundary conditions only on the outer edges of the geometry, while internally (i.e., in the bulk), the results are determined from the governing equations and boundary conditions (details on the numerical simulations can be found in the Supporting Information[22]). **Figure 3**(a)-(b) shows the electric potential distribution, $\phi$ , in a 2D and 3D system, respectively. The black lines, which denote the streamlines of the electric field ($\boldsymbol{E} = -\nabla\phi$), show that field lines focus (and defocus) at the microchannel-nanochannel interface. The green lines are internal lines of symmetry, calculated from numerical simulations, that demonstrate that adjacent unit-cells do not exchange electric field flux between themselves.

### 2.2. Parallel circuit

The independence of each unit-cell suggests that the equivalent electrical circuit of an array of identical unit-cells can be described as an array of $N$ unit-cells connected in parallel. In the nanofluidic system, this means that the unit-cells act as independent and identical paths for ionic fluxes across a potential drop, $V$. The total resistance is given by

$$R_{\text{total}} = \frac{R_{\text{unit-cell}}}{N}, \qquad (1)$$

where, $R_{\text{unit-cell}}$ [Eq. (2), **Figure 2**(d)], is the resistance of a single unit-cell system (discussed below). In the remainder of this paper, we will demonstrate that the predicted $R_{\text{total}} \sim N^{-1}$ scaling, holds both numerically and experimentally (Sec. 3). To that end, in the following sub-section, we will discuss $R_{\text{unit-cell}}$, while in Sec. 4.2, we will compare our model with other existing models that predict different scaling with $N$. We already note here, what is discussed throughout this work, and in particular in Sec. 4.2, is that in a multichannel system, it is unit-cells, rather than individual nanochannels that are mutually independent. If it were nanochannel resistances, $R_{\text{nano}}$, that were independent, then we would have $R_{\text{total}} = R_{\text{nano}}/N$. However, as we will emphasize throughout this work since the microchannels' contributions are non-negligible, this is not the case.

### 2.3. The unit-cell resistance

The unit-cell [**Figure 2**(b)] is a single nanochannel system where two cuboidal microchannels are bridged by one nanochannel of arbitrary cross-sectional geometry. The microchannels have a height, $H$, width, $W$, and length, $L$. Here, for the sake of simplicity, we have depicted the nanochannel as a cylinder of radius, $a$, and length, $l$. In general, it can be of any perimeter, $P_{\text{nano}}$, and cross-section area, $S_{\text{nano}}$ so long as the aspect ratio $l/S_{\text{nano}}^{1/2}$ is large ($l/S_{\text{nano}}^{1/2} \gg 1$).

The resistance of the above-described system, $R_{\text{unit-cell}}$, when the nanochannel has a surface charge density, $\sigma_s$, was recently derived in our previous work [23] with two differences: a) The nanochannel was cuboidal. b) The nanochannel was centered at the center of the unit-cell. Here we alleviate both assumptions and consider a more general scenario.

The outcome of removing these assumptions is discussed further below. Thus, it is essential to emphasize that the novelty of this work is not in $R_{\text{unit-cell}}$ but rather in its modification (Sec. 2.3.1) and primarily with the application of $R_{\text{unit-cell}}$ within the framework of the parallel circuit [Eq. (1)].

For the case of a charged nanochannel with a surface charge density, $\sigma_s$, it has been shown from the exact analytical solution[23] of the Poisson-Nernst-Planck (PNP) equations that the total Ohmic resistance and the transport number of the unit-cell are given by

$$R_{\text{unit-cell}} = (2\tau - 1)\frac{R_{\text{nano}} c_0}{\Sigma_s} + \left[1 + (2\tau - 1)\sqrt{4\frac{c_0^2}{{\Sigma_s}^2} + 1}\right]\frac{\Sigma R_{\text{micro}}}{2}, \qquad (2)$$

$$\tau = \frac{1}{2} + \frac{\Sigma_s}{2c_0}\left(\sqrt{4 + \frac{\Sigma_s^2}{c_0^2}} + 2\frac{\Sigma R_{\text{micro}}}{R_{\text{nano}}}\right)^{-1}, \qquad (3)$$

where the resistances of the nanochannel, $R_{\text{nano}}$, and microchannels, $R_{\text{micro}}$, are given by

$$R_{\text{nano}} = \rho_{\text{res}}\frac{l}{S_{\text{nano}}}, R_{\text{micro}} = \rho_{\text{res}}\frac{L}{S_{\text{micro}}},$$
$$\Sigma R_{\text{micro}} = 2(R_{\text{micro}} + R_{\text{FF}}), \qquad (4)$$

and the average excess counterion concentration at every cross-section of the nanochannel is

$$\Sigma_s = -\frac{\sigma_s}{F}\frac{P_{\text{nano}}}{S_{\text{nano}}}. \qquad (5)$$

Here the resistivity is given by $\rho_{\text{res}} = \Re T/(F^2 D c_0)$ where $\Re$ is the universal gas constant, $T$ is the (absolute) temperature, $F$ is the Faraday constant, $D$ is the diffusion coefficient of the ions, and $c_0$ is the bulk concentration of the ions in the reservoirs. Note that here we have assumed that the electrolyte is KCl which can be assumed to be a symmetric binary electrolyte- i.e., both ionic species have the same valences ($z_\pm = \pm 1$) with equal diffusion coefficients ($D_\pm = D$). It is immediate from Eq. (5) that if the surface charge density is negative, $\sigma_s < 0$, the excess concentration is positive, $\Sigma_s > 0$, and vice-versa.

Both the nanochannel and microchannel resistances scale with their lengths divided by their respective cross-sectional areas. For the microchannel, this area is always $S_{\text{micro}} = HW$. In contrast, the nanochannel area $S_{\text{nano}}$ depends on the details of the nanochannel cross-section itself. In our previous work[23], we considered a long cuboidal nanochannel such that $S_{\text{nano}} = hw$. However, in this work, we can also consider a long cylindrical nanochannel of radius $a_{cylinder}$ such that $S_{\text{nano}} = \pi a_{cylinder}^2$. In fact, we can consider a nanochannel of arbitrary shape so long as it satisfies $l/S_{\text{nano}}^{1/2} \gg 1$.

The primary difficulty in varying the geometry from a cuboidal geometry to another geometry can be associated with the second term of the total resistance due to the microchannels, $\Sigma R_{\text{micro}}$, which also include the resistances associated with the microchannel-nanochannel interfaces. We denote this resistance $R_{\text{FF}}$ and discuss this term thoroughly in the following subsection (Sec. 2.3.1). In Sec. 2.3.2, we will discuss the significant role the transport number has in determining the response of the system. We will demonstrate that the total resistance cannot always be represented by a simple electrical circuit. In fact, we will show that the Thevenin method of presupposing a circuit form is generally inapplicable to nanofluidic systems. Finally, note that the factor 2 in Eq. (4) represents the contribution of both (inlet and outlet) microchannels.

### *2.3.1. Field-focusing resistance*

From as far back as 1975, Hall[24] showed that the entrance effects to a circular nanochannel contributed to the nanochannel in a non-negligible manner. One recognizes $R_{\text{FF}}$ to be $R_{\text{access}} = \rho_{\text{res}}/(4a_{\text{cylinder}})$ which is the classical access resistance solution [24–26]. However, $R_{\text{access}}$ makes three limiting assumptions. First, there is a single pore. Second, this pore is an infinite medium/domain and does not interact with walls (or other pores). Third, the pore is circular. However, in most experimental systems[27–29], none of these assumptions hold. To overcome these difficulties, we have introduced the generalizable access resistance, which we term the field-focusing resistance, $R_{\text{FF}}$.[23,30,31] The field-focusing resistance makes none of Hall's assumptions.

To the best of our knowledge, there are only two known solutions for $R_{\text{FF}}$. The first is for the classical access resistance, $R_{\text{access}}$ [**Figure 4** (a)]. The second, derived by us in a set of past works[23,30,31], investigated cuboidal nanochannels interfacing with cuboidal microchannels [**Figure 4** (b)]. In fact, Eq. (2) was explicitly derived for cuboidal nanochannels. One of the novelties of this work is to demonstrate that Eq. (2) holds for nanochannels of arbitrary cross-sectional geometries and to provide an expression for $R_{\text{FF}}$ that can be consistently calculated for any geometry, confined within a unit-cell (satisfies $\boldsymbol{j} \cdot \boldsymbol{n} = 0$), as shown in **Figure 4**. The general expression for $R_{\text{FF}}$ can be found in the Supporting Information[22], while **Table S2** provides the expression for $R_{\text{FF}}$ for the widely investigated cases of cuboidal and cylindrical nanochannels.

Three last comments regarding $R_{\text{FF}}$ are essential to understanding the modeling in this work. First, $R_{\text{FF}}$ is independent of the ratio $\Sigma_s/c_0$. Second, as will be discussed shortly, $R_{\text{FF}}$ is derived for the unit-cell that satisfies $\boldsymbol{j} \cdot \boldsymbol{n} = 0$. As adjacent unit cells do not exchange flux between themselves, $R_{\text{FF}}$ is independent of the total number $N$ of nanochannels in the array – this result differs from other approaches (see Discussion – Sec. 4.2). Third, since $R_{\text{FF}}$ no longer assumes that the nanochannel is embedded in an infinite medium, one should longer assume that the effects of the microchannels are negligible. Thus, an-other difference between $R_{\mathrm{FF}}$ and $R_{\mathrm{access}}$ is that $R_{\mathrm{micro}}$ is no longer negligible in $\Sigma R_{\mathrm{micro}}$. Here we consider the most general, realistic scenario and retain $R_{\mathrm{micro}}$.

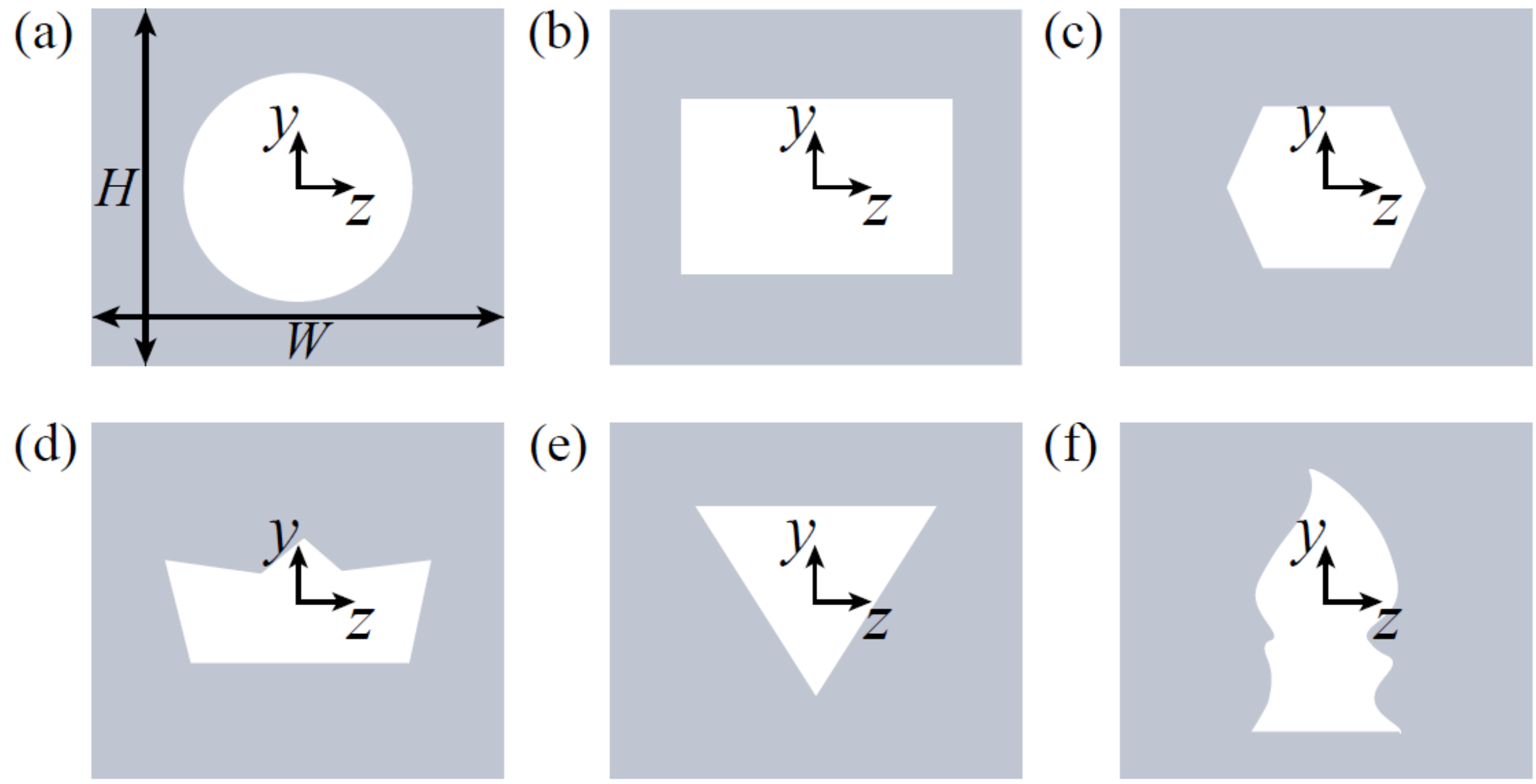

**Figure 4** The new expression for $R_{\mathrm{FF}}$ derived in this work (see Supplemental Information[22]) can be used to calculate the field focusing resistance for nanochannels of arbitrary cross-section at the nanochannel-microchannel interfaces. This includes (a) a centered cylindrical nanochannel[19,32], (b) a cuboidal channel[7,33], (c) a hexagon nanochannel, (d) a crown-shaped nanochannel[27], (e) a triangular channel[28], and (f) truly arbitrary cross-sections (here we have used the silhouette of our university logo).

#### *2.3.2. Equivalent circuit of a nanochannel*

It is important to make several comments regarding Eq. (2). First, in contrast to other models that will be discussed in Sec. 4.2, Eq. (2) is derived from the PNP equations with minimal assumptions[23]. Thus, $R_{\text{unit-cell}}$ has precedence over other models that are not derived rigorously. Second, $R_{\text{unit-cell}}$ [Eq. (2)], through the transport number, $\tau$, depends non-linearly on the characteristic resistances $R_{\mathrm{nano}}$ and $\Sigma R_{\mathrm{micro}}$ such that in general $R_{\text{unit-cell}}$ [Eq. (2)] cannot be represented as a simple series circuit. This is further complicated with $R_{\text{unit-cell}}$'s dependence on the ratio $\Sigma_s/c_0$ (Sec. 4.2)

These comments demonstrate that in the most general scenario, Thevenin's theorem, which presupposes an existing equivalent electrical circuit, cannot be used to represent $R_{\text{unit-cell}}$ with simple resistors connected in series. However, if $R_{\text{unit-cell}}$ is known precisely and the concept of unit-cells holds, one can use Thevenin's theorem to represent the total resistance of an array as a parallel circuit (Sec. 2.2).

#### *2.3.3. Series circuit*

Nonetheless, without retracting from the general comments above, it is useful to demonstrate that in the two distinct limits $\Sigma_s/c_0 \ll 1$ and $\Sigma_s/c_0 \gg 1$, Eq. (2) can be described as a set of serially connected resistances [Figure 2(d)] whereby the coefficients multiplying $R_{\mathrm{nano}}$ and $\Sigma R_{\mathrm{micro}}$ are no longer explicitly dependent on $\tau$ and have a very simple dependence on $\Sigma_s/c_0$.

*Vanishing selectivity:* The limit of $\Sigma_s/c_0 \ll 1$ corresponds to the case that the excess counterion concentration within the nanochannel is negligible compared to the bulk concentration. This case is commonly referred to as the limit of vanishing selectivity, where the nanochannel does not filter any of the coions, and the electric response of the system depends only on bulk properties (i.e., this is the bulk response of the system and is independent of the surface charge density). In this scenario, it can be shown[23] that $\tau \cong \frac{1}{2}$ such that the unit-cell resistance is given by

$$R_{\text{unit-cell}}^{(\text{vanishing})} = \frac{1}{2}\rho_{\mathrm{res}}(R_{\mathrm{nano}} + \Sigma R_{\mathrm{micro}}). \tag{6}$$

Here, the resistance scales with the resistivity and hence scales inversely with the concentration, $R_{\text{unit-cell}}^{(\text{vanishing})} \sim \rho_{\mathrm{res}} \sim c_0^{-1}$.

*Ideal selectivity:* At the limit of $\Sigma_s/c_0 \gg 1$, it can be shown that $\tau = 1$. This is the all-important limit of ideal selectivity where the nanochannel exhibits perfect coion exclusion. Ideal selectivity lies at the heart of ED and RED systems. In this scenario, the unit-cell resistance is given by

$$R_{\text{unit-cell}}^{(\text{ideal})} = \rho_{\mathrm{res}}\left[\frac{R_{\mathrm{nano}}}{(\Sigma_s/c_0)} + \Sigma R_{\mathrm{micro}}\right]. \tag{7}$$

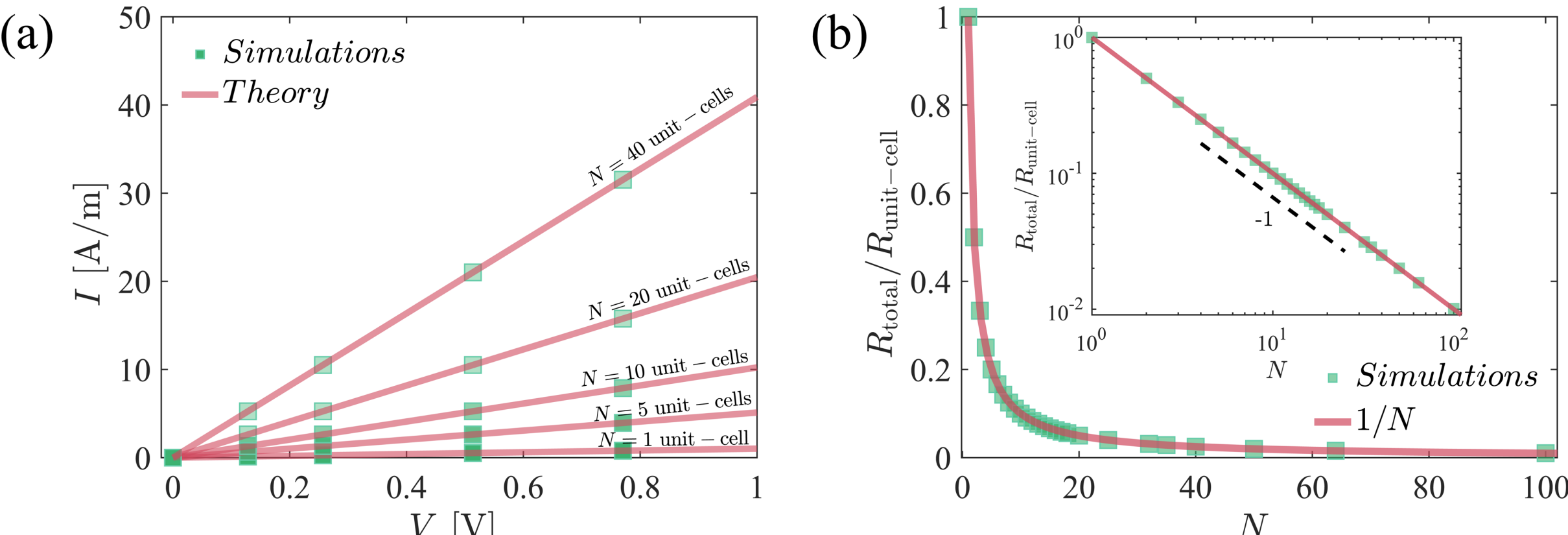

**Figure 5** (a) The current-voltage response $(I - V)$ of single and multichannel 2D arrays [**Figure 3**(a)]. (b) The ratio of the total resistance to the unit-cell resistance, $R_{\text{total}}/R_{\text{unit-cell}}$, versus the number of unit-cells, $N$, follows the expected $1/N$ scaling of Eq. (1). (inset) A $\log_{10}-\log_{10}$ plot of $R_{\text{total}}/R_{\text{unit-cell}}$ versus $N$. The dashed line has a slope of -1. We note there that the numerical simulations were conducted in non-dimensional format – this is discussed in the Supporting Information[22] – however, here we have presented the result after the voltage was dimensionalized by the thermal temperature at room temperature, $\phi_{\text{th}}(T = 298K)$, and the current was dimensionalized by $\rho_{res}/h$. For a KCl electrolyte at room temperature, concentration of $c_0 = 1\ [M]$ and diffusion coefficient of $D = 2 \times 10^{-9}\ [m^2/s]$, $\rho_{res} = 0.1331\ [\Omega m]$ and the characteristic length is $h = 10$ [nm]. The geometric properties can be found in **Table S1** of the Supporting Information[22].

*Differences between ideal and vanishing selectivity:* The three substantial differences between Eqs. (6) and (7) are the following. First, in Eq. (6) the factor $\frac{1}{2}$ is due to the equal but oppositely directed transport of coions and counterions. Twice the number of charge carriers in the conducting medium halves the resistance. Second, in Eq. (7) the nanochannel resistance is modified by the $\Sigma_s/c_0$ term. This term captures the nanochannel's ability to ideally exclude coions. This term also leads to the third significant difference. For a nanochannel system, at least one of its characteristic lengths ($h$, $w$, or both) is substantially smaller than the length, $l$, such that $R_{\text{nano}}$ dominates $R_{\text{unit-cell}}$. However, in Eq. (7), $R_{\text{nano}}$ is multiplied by $c_0/\Sigma_s$ such that for a given geometry, at sufficiently low concentrations ($c_0 \to 0$ leading to $\Sigma_s/c_0 \gg 1$), $R_{\text{nano}}c_0/\Sigma_s$ is no longer the dominant term[23,31]. Then the electrical response is determined by $\Sigma R_{\text{micro}}$. See Ref. [23] for a detailed discussion regarding the differences between $R_{\text{unit-cell}}$ [Eq. (2)] to $R_{\text{unit-cell}}^{(\text{vanishing})}$ [Eq. (6)] and $R_{\text{unit-cell}}^{(\text{ideal})}$ [Eq. (7)].

## 3. Results

In this section, we will present numerical (Sec. 3.1)and experimental (Sec. 3.2) evidence that supports the prediction that $R_{\text{total}} \sim N^{-1}$ [Eq. (1)].

### 3.1. Parallel circuits in 2D

To ensure that our 2D numerical simulations correspond to the theoretical analysis, we calculated the electric current, $I$, for a given potential drop, $V$. This is the current-voltage response $(I - V)$. The markers in **Figure 5**(a) show the numerically computed $I - V$ of single- and multi- unit-cell systems. Our theoretical lines are the response of the single unit cell current multiplied by the number of channels, $N$. The perfect correspondence confirms the prediction of Eq. (1). **Figure 5**(b) shows the total resistance $R_{\text{total}}$ has the predicted $N^{-1}$ dependence. In the Supporting Information[22], we repeat this analysis for 3D arrays of cuboidal and cylindrical nanochannels – the results remain unchanged.

### 3.2. Experiments on 3D arrays

We now demonstrate, using the experimental work of Ref.[33], that nanochannel arrays follow the $N^{-1}$ scaling of Eq. (1). To that end, we use their data and further extend their analysis. It is necessary to discuss the difference in how the geometry is here relative to how it is defined in Ref.[33].

**Figure 6**(a) is a schematic of the multichannel (line-) array system used in Ref.[33], segmented to highlight the corresponding unit-cells. The geometry of the unit-cells in **Figure 6**(a) is, in fact, half that of the general unit-cell considered in our above analysis [**Figure 2**(b)]. As a result, the analytical expression for $R_{\text{FF}}$ varies. Since we provide the final solution here in terms of $R_{\text{FF}}$ our approach means unaltered once $R_{\text{FF}}$ is calculated correctly for the suitable unit-cell geometry. The differences between the expressions for the "full-cell" and "half-cell", and how to identify the unit-cell, are further discussed in the Supporting Information[22].

Also, it should be noted that Ref.[33] focused on the effects of interchannel spacing and the effects of electroconvection on the overall electrical response of multichannel systems. Also, it should be noted that Ref.[33] focused on the effects of interchannel spacing and the effects of electroconvection on the overall ele-

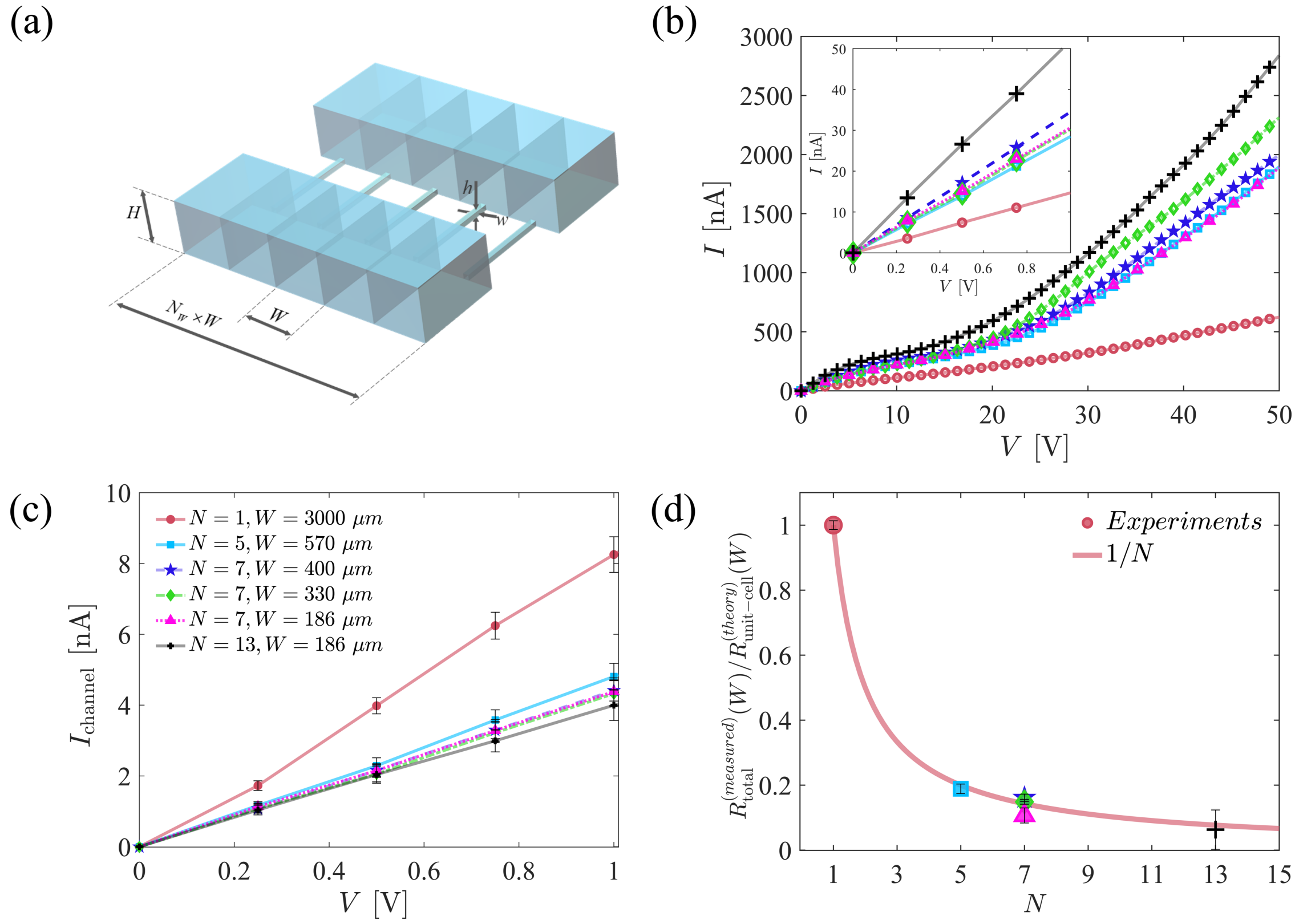

**Figure 6** (a) Schematic of the experimental setup of Ref.[33] of nanochannel arrays of varying $N$ and varying interchannel spacing, $W$. Note that these channels are half-cells relative to the geometry shown in Figure 2(a)-(b) [see main text for additional discussions about half-cells]. (b) The current-voltage $(I - V)$ response. (Inset) Zoom up the low-voltage Ohmic response. (c) The current-per-channel-voltage response $(I_{\text{channel}} - V)$. (d) The ratio $R_{\text{total}}^{(\text{measured})}(W)/R_{\text{unit-cell}}^{(\text{theory})}(W)$ [Eq. (1)] versus $N$.

ctrical response of multichannel systems. Therefore, they were not interested in the Ohmic response of the devices and the dependence of $R_{\text{total}}$ on $N$. Here, we extend the analysis of their data to delineate the Ohmic response and verify Eq. (1).

The experiments of Ref.[33] were conducted on several multichannel array systems with the same height, $H$, and length, $L$, but with a varying number of identical nanochannels, $N$, and varying interchannel spacing, $W$. Note that the interchannel spacing, $W$ is equal to the width of the corresponding unit-cell. The array spacing and the number of channels are given in the legend of **Figure 6**(c) and in **Table 1**.

For each configuration, numerous $I - V$ curves were measured. The mean $I - V$ are plotted in **Figure 6**(b), where the inset focuses on the low-voltage Ohmic response. **Figure 6**(b) demonstrates the expected result that increasing $N$ leads to an increase in the total current. This makes physical sense – as $N$ increases, $R_{\text{total}}$ decreases, as the total area through which the flux is transported increases correspondingly. **Figure 6**(c) shows the current per channel, $I_{\text{channel}} = I/N$ from which it can be observed that channels with larger $W$ have larger currents – this, too, is consistent with the corresponding reduction of $R_{\text{micro}}$ and $R_{\text{FF}}$.

Using the data from Ref.[33], we calculate $R_{\text{total}}^{(\text{measured})}(W) = V/I$. Then, using Eq. (7), we isolate $\Sigma_s$ from the single-channel system $(N = 1)$ and find that $\Sigma_s = 5.05[\text{mol/m}^3]$ (such that $\Sigma_s/c_0 = 168.3$) and the calculated surface charge density is $\sigma_s =$ -0.043[$\text{C/m}^2$]. This value extracted from the single-channel system is then used to calculate the unit-cell resistances of all the other array systems, as predicted by Eq. (7) as a function of the interchannel spacing, $R_{\text{unit-cell}}^{(\text{theory})}(W)$. In **Figure 6**(d), we calculate the ratio $R_{\text{total}}^{(\text{measured})}(W)/R_{\text{unit-cell}}^{(\text{theory})}(W)$. We demonstrate that this ratio varies as $N^{-1}$, as predicted by Eq (1).

## 4. Discussion and future directions

In the following section, we will discuss the outcomes of the results presented in this work and their relation to several recent works that have covered a wide range of topics related to the physics and applications of nanochannel arrays.

### 4.1. Additional experimental evidence

Esfandiar et al.[34] considered an array of $N =$ 200 channels in a parallel, line array configuration [similar to that of **Figure 6**(a)]. Esfandiar et al.[34] upscaled from a single channel system to a 200-

**Table 1** Experimental parameters of the systems from Ref.[33]. All systems have the same nanochannel height, $h = 178$ [nm], nanochannel width, $w = 92$ [μm], nanochannel length, $l = 350$ [μm], microchannel height, $H = 48$ [μm], and microchannel length, $L = 2$ [mm]. The number of channels in each system was varied, as well as the unit-cell width $W$. All experiments were conducted with a KCl solution of concentration $c_0 = 30$ [μM], at room temperature (298 [$K$]), such that the resistivity is $\rho_{res} = 4436$ [Ωm]. In general, the nanochannel resistance contribution $R_{\text{nano}}/(\Sigma_s/c_0) = 0.56$ [$G\Omega$] is smaller than the microchannel contribution $\Sigma R_{\text{micro}}$.

| $N$ | $W\,[\mu m]$ | $R_{\text{micro}}\,[M\Omega]$ | $R_{\text{FF}}\,[M\Omega]$ | $\Sigma R_{\text{micro}}\,[G\Omega]$ |
|---|---|---|---|---|
| 1 | 3000 | 61.61 | 264.98 | 0.65 |
| 5 | 570 | 324.26 | 216.26 | 1.08 |
| 7 | 400 | 462.07 | 205.95 | 1.33 |
| 7 | 330 | 560.08 | 200.39 | 1.52 |
| 7 | 186 | 993.70 | 184.21 | 2.36 |
| 13 | 186 | 993.70 | 184.21 | 2.35 |

channel system to increase the sensitivity of their measurements. In their Figure S3, they show that at high concentrations (i.e., vanishing selectivity), the conductance per channel of the 200-channel configuration was equal to that of the single channel. This is unsurprising since, at this limit, the nanochannel resistance dominates the unit-cell response, and thus the multiplication of the unit-cell conductance by $N = 200$ is rather intuitive. At low concentrations (i.e., ideal selectivity), the conductance per channel of the 200-channel configuration was two orders of magnitude smaller than that of a single channel. This, too, is unsurprising since, at this limit, the nanochannel is no longer the dominant resistance. In fact, for such a highly packed system, the effects of $R_{\text{micro}}$ and $R_{\text{FF}}$ are prevalent, especially if $R_{\text{micro,200}} \sim 200 R_{\text{micro,1}}$.

In this work, we have considered the much simpler (ordered) array comprised of identical unit-cells. While future works should consider how to extend our approach to systems with multiple unit-cell types, it is important to note that the universality of our approach of electrical circuit modeling still holds. For example, recently, Lucas and Siwy[19] explored the possibility of designing ionic circuits based on nanochannel arrays. They simulated four different array configurations ($3 \times 3, 2 \times 3, 1 \times 3$ and $1 \times 1$) of long cylindrical nanochannels. Thus, depending on the configuration, one can identify several distinct unit-cell types. For example, in their $3 \times 3$ configuration shown in the inset of **Figure 7**, there are three distinct unit-cells: the central unit-cell, the corner unit-cells, and the center-face unit-cells. Accordingly, in their Figure 2(b)[19], which we reproduce here[35,36] in **Figure 7**, one can observe that the current per channel depends on the esoteric unit-cells within each system, indicating that the concept of unit-cells also holds for arrays of non-identical microchannels. However, identifying the boundary of each unit-cell a priori for a general array configuration is not a trivial problem and is left for future work.

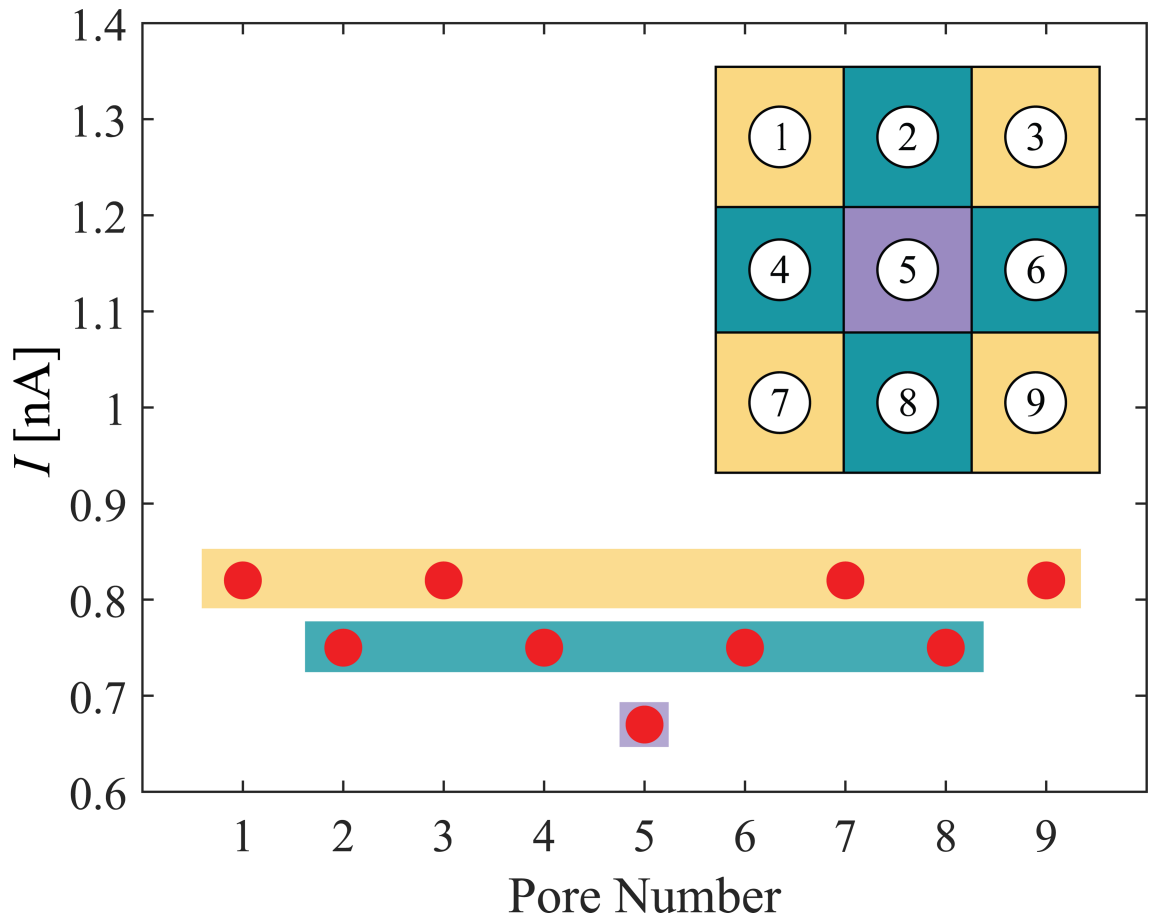

**Figure 7** The calculated current for each channel shown in the inset. It can be observed that there are three different types of unit-cells and three different values of the current. Each unit-cell has a different resistance and, hence, a different current. Data reproduced from Ref.[19] (American Chemical Society 2020, available under CC-BY-NC-ND license).

While our work focuses on arrays of long nanochannels, the qualitative nature of the dependence of $R_{\text{FF}}$ should hold for arrays of short nanochannels (i.e., when the diameter and pore length are of the same order of magnitude). Yazda et al.[37] recently fabricated an array within a hexagonal boron nitride/silicon nitride membrane. They measured the dependence of the generated osmotic current on the pore spacing and reported that the current increased with interchannel spacing. This, too, can be discerned with the understanding that highly isolated channels have larger currents. Such a result is consistent with this work as well (discussed further in Sec. 4.3).

## 4.2. Equivalent circuits in nanofluidics

### *4.2.1. The fallibility of Thevenin's theorem*

In nanofluidics, it is common to use Thevenin's theorem, which suggests that the equivalent electrical circuit of the system can be expressed simply by adding various resistances. As already indicated in Sec. 2.3.2, such a statement is problematic. We will demonstrate this using the most common misconception in nanofluidics. To demonstrate the common misconception, rather than discussing the Ohmic resistance $R_{\text{Ohmic}}$, we will discuss the Ohmic conductance, which is reciprocal to the resistance ($G = R^{-1}$).

At high concentrations ($\Sigma_s/c_0 \ll 1$), the conductance is determined by the bulk concentration, $G_{\text{bulk}} = 2S_{\text{nano}}/(\rho_{\text{res}} l)$, while at low concentrations ($\Sigma_s/c_0 \gg 1$), the conductance is concentration-independent, $G_{\text{surface}} = \Sigma_s S_{\text{nano}}/(\rho_{\text{res}} c_0 l) \sim c_0^0$ (often termed 'surface conductance'). It is common to suggest that the sum of $G_{\text{bulk}}$ and $G_{\text{surface}}$ (i.e., their superposition)

$$G_{\text{super-position}} = G_{\text{bulk}} + G_{\text{surface}} = \frac{S_{\text{nano}}}{l}\left(\frac{2}{\rho_{\text{res}}} + \frac{\Sigma_s}{\rho_{\text{res}} c_0}\right), \quad (8)$$

captures the total conductance. The popularity of Eq. (8) is due to the simple handwaving explanation associated with it. This is tantamount to stating that ions are transported parallelly through either the bulk conductance or the surface conductance. Since $G = R^{-1}$, Eq. (8) can be rewritten

$$\frac{1}{R_{\text{super-position}}} = \frac{1}{R_{\text{bulk}}} + \frac{1}{R_{\text{surface}}}, \quad (9)$$

which represents that surface and bulk resistances are connected in parallel. However, as we have demonstrated in several past works[23,31], Eq. (8) [and Eq. (9)] is incorrect. Before briefly explaining the origin of this common misconception and the origin of the error, it is worthwhile to present the correct solution. Consider Eq. (2) under the assumption of negligible microchannel effects ($\Sigma R_{\text{micro}} = 0$). Equation (2) takes the renowned form [7,38–40].

$$G_{\text{unit-cell}}^{(\Sigma R_{\text{micro}}=0)} = \frac{1}{R_{\text{unit-cell}}^{(\Sigma R_{\text{micro}}=0)}} = \sqrt{4 + \frac{\Sigma_s^2}{c_0^2}}\,\frac{S_{\text{nano}}}{\rho_{\text{res}} l}. \quad (10)$$

Observe that at the two limits of $\Sigma_s/c_0 \ll 1$ and $\Sigma_s/c_0 \gg 1$, Eqs. (8) and (10) are identical while they are different for $\Sigma_s/c_0 \sim 1$. However, Eq. (8) was artificially constructed to yield the correct high and low concentration limits. Further, there is no known self-consistent derivation for Eq. (8) – it is primarily based on empirical reasoning. In contrast, Eq. (10) represents an exact solution of the PNP equations and can be derived using two different approaches (with[23] and without microchannels[41]). Further, numerical simulations have shown that Eq. (10) captures the $\Sigma_s/c_0 \sim 1$ response accurately[23]. (see Ref.[23] for a thorough discussion on other shortcomings of the superposition model).

Since Eq. (10) is the correct expression, it is important to observe that this expression is not given by a simple electrical circuit comprised of two electrical resistances, as suggested by Eq. (8). In fact, the failure of the superposition model is of immense importance. It demonstrates that using Thevenin's method of presupposing an electrical circuit is incorrect for the simplest imaginable scenario of a single nanochannel-only system. Hence, any further use of Thevenin's theorem should be made with careful supervision.

We repeat our precautionary comment regarding Eq. (2) concerning parallel circuits. While Eq. (2) can be simplified at the two limits of $\Sigma_s/c_0 \ll 1$ and $\Sigma_s/c_0 \gg 1$ [Eqs. (6) and (7), respectively] so that these resultant equations can be simplified to be a series circuit, in general, this is not the case. This is because of the dependence of $R_{\text{unit-cell}}$ on the transport number, $\tau$ (see Sec. 2.3.2). However, because $R_{\text{unit-cell}}$ is derived under the assumption of zero-flux ($\boldsymbol{j} \cdot \boldsymbol{n} = 0$), so long as $\boldsymbol{j} \cdot \boldsymbol{n} = 0$ holds, Eq. (1) holds.

### *4.2.2. Scaling of $R_{\text{total}}$ on $N$*

We shall now present alternative models that propose alternative scaling of the total resistance with $N$.

Gadaleta et al.[32] suggested, in the notation of this work, that as $N \to \infty$ the total conductance, which is reciprocal to the resistance ($R_{\text{total}} = G_{\text{total}}^{-1}$), goes to zero such that $\lim_{N\to\infty}(G_{\text{total}}/N) \to 0$. In the notation of this work, the high concentration (i.e., vanishing selectivity) conductance per Gadaleta et al.[32] is

$$G_N = N G_{\text{unit-cell}} = \frac{N}{\rho_{\text{res}}}\left(\frac{4l}{\pi a_{\text{cylinder}}^2} + \frac{1}{2a_{\text{eff}}}\right)^{-1}. \quad (11)$$

The difference between our work and theirs originates in how we identify the unit-cell and how we enforce the no-flux condition, as well as determining the relative contribution of all terms:

- They always neglect $R_{\text{micro}}$ whereas we keep this term.
- They modify Hall's access resistance solution, $R_{\text{access}}$, to account for the finite distance between two pores. To account for the numerous pores, Gadaleta et al.[32] modified the single pore access resistance [i.e., the second term in the brackets of Eq. (11)] such that it depends on an 'effective radius' and pore spacing, H, whereby $a_{\text{eff}}/a_{\text{cylinder}} = [1 + \gamma_N a_{\text{cylinder}}/(H)]^{-1}$. They state that $\gamma_N$ is a "global factor accounting for the geometry of the network", and they provide different expressions for $\gamma_N$

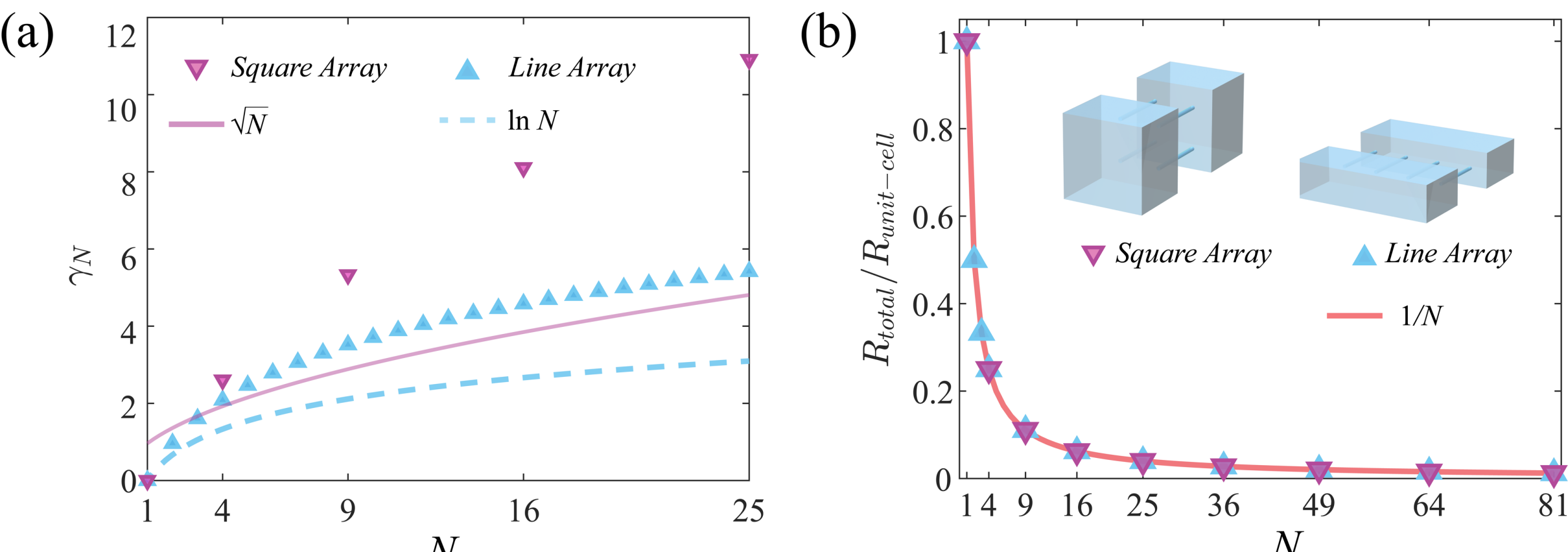

**Figure 8** (a) Comparison of the parameter $\gamma_N$, given by Eqs. (15) and (18) of Ref.[32] for a line array of nanopores and a square array of nanopores, respectively. The solid lines present the proposed $\sqrt{N}$ and $\ln N$ scaling of Ref. [32]. (b) The normalized total resistance, $R_{\text{total}}/R_{\text{unit-cell}}$ for a line array and square array that have $N$ channels within square unit-cells ($W = H$). The channels are highly packed with $H/(2a) = 2$, $L = 200a$, and $l = a/2$.

when it is a line array and a square array [their Eqs. (15)-(19)]. They state that $\gamma_{N,\text{line-array}} \sim \ln N$ [their Eq. (16)] and $\gamma_{N,\text{square-array}} \sim N^{1/2}$ [their Eq. (19)].

- Accordingly, in their description[32], $G_{\text{unit-cell}}$ is implicitly dependent on $N$ whereas in this work $R_{\text{unit-cell}}$ is independent of $N$. This difference can be attributed to how no-flux is enforced. We require explicitly require $\boldsymbol{j} \cdot \boldsymbol{n} = 0$. at the edges of our unit-cells, whereas they[32] used an image mirroring approach to account for the various interactions. However, their approach was limited to accounting only for first-order interactions between all pores.
- Within their proposed framework, $\lim_{N\to\infty} (G_{\text{total}}/N) \to 0$. Such a suggestion is in contradiction with Eq. (1), which yields $G_{\text{total}}/N = (R_{\text{total}}N)^{-1} = R_{\text{unit-cell}}^{-1}$.

We have several comments regarding their model. First, if one calculates $\gamma_N$ as prescribed by their Eq. (15) and Eq. (18), it can be observed that neither has the correct proposed scaling [$\ln N$, $N^{1/2}$- see **Figure 8** (a)]. Second, 3D arrays of the same number of $N$ total channels can be constructed in several ways. For example, one can construct a single-line array of $1 \times N$, or double-lined array of $2 \times (\frac{1}{2}N)$, or a square array of $\sqrt{N} \times \sqrt{N}$ (when $N$ is a square number). One can then ask if the scaling of $R_{\text{total}}$ with $N$ changes. We will show that it does not. To that end, we simulate a system of line arrays and square arrays similar to the one considered in the experiments of Ref.[32], where the cylindrical nanochannels of radius $a$ were highly packed $H/(2a) = 2$[inset of **Figure 8** (b)]. In both scenarios, we considered cubical microchannels ($W = H$). For this geometry, the ratio $R_{\text{micro}}/R_{\text{FF}} = 151$ such that $R_{\text{micro}}$, and not $R_{\text{FF}}$, dominates $\Sigma R_{\text{micro}}$. **Figure 8** (b) shows that line arrays and square arrays of $N$ channels have the predicted $N^{-1}$ scaling [Eq. (1)] and are independent of the geometric configuration.

Another approach uses impedance modeling principles whereby a circuit that fits the current-voltage response is constructed (similar to Thevenin's theorem). For example, consider Figure 1 of Morikawa et al.[42] here an electrical circuit is prescribed for a system of $N$ nanochannels under the presumption that all the nanochannels operate as current sources connected in parallel and are thereafter connected to the microchannel resistances (the effects of field focusing are not considered, but the effects of a load resistance, $R_{\text{load}}$, is considered). In the notation of this work, the electrical circuit can be described as $R_{\text{total}} = (R_{\text{nano}}/N + 2R_{\text{micro}} + R_{\text{load}})$. Their model/analysis has two interesting peculiarities. First, they assume that the entire model is dominated by an infinitely large $R_{\text{load}}$ and the effects of the nanochannels and microchannels are negligible. This is somewhat surprising as this results in an electrical current that is independent of the nanochannel and microchannel geometries. Another peculiarity is that at the limit of an infinite number of channels, $N \to \infty$, the resistance is once more independent of the nanochannel geometry. In contrast, our Eqs. (1) and (2) do not exhibit such a peculiar behavior.

In this section, we have further demonstrated the fallibility of Thevenin's theorem. We disagree with the use of Thevenin's approach and advocate the use of a more cautious approach that utilizes less handwaving explanations and rather utilize solving the PNP equations.

### 4.3. Scaling up the electrical current

In this section we address a conceptual problem that needs to be settled concerning the parallelization of single-channel systems used in desalination and energy harvesting systems. Namely, we will demonstrate that by varying the geometry, one can optimize either the total current or the current density – but not both. Siria et al.[5] suggested that the single pore RED systems can reach current densities up to 0.8[kWhm$^{-3}$] and that one could reach substantial energy yields via parallelization. In contrast, Wang et al.[43] argued that parallelization would not yield the promised large fluxes due to the appearance of concentration polarization associated with multichannel systems. It should be stated that the analysis of both works is presented from different perspectives – current densities versus total currents – and that both are correct. Our model and analysis, given below, can resolve this dichotomy by providing a complete picture.

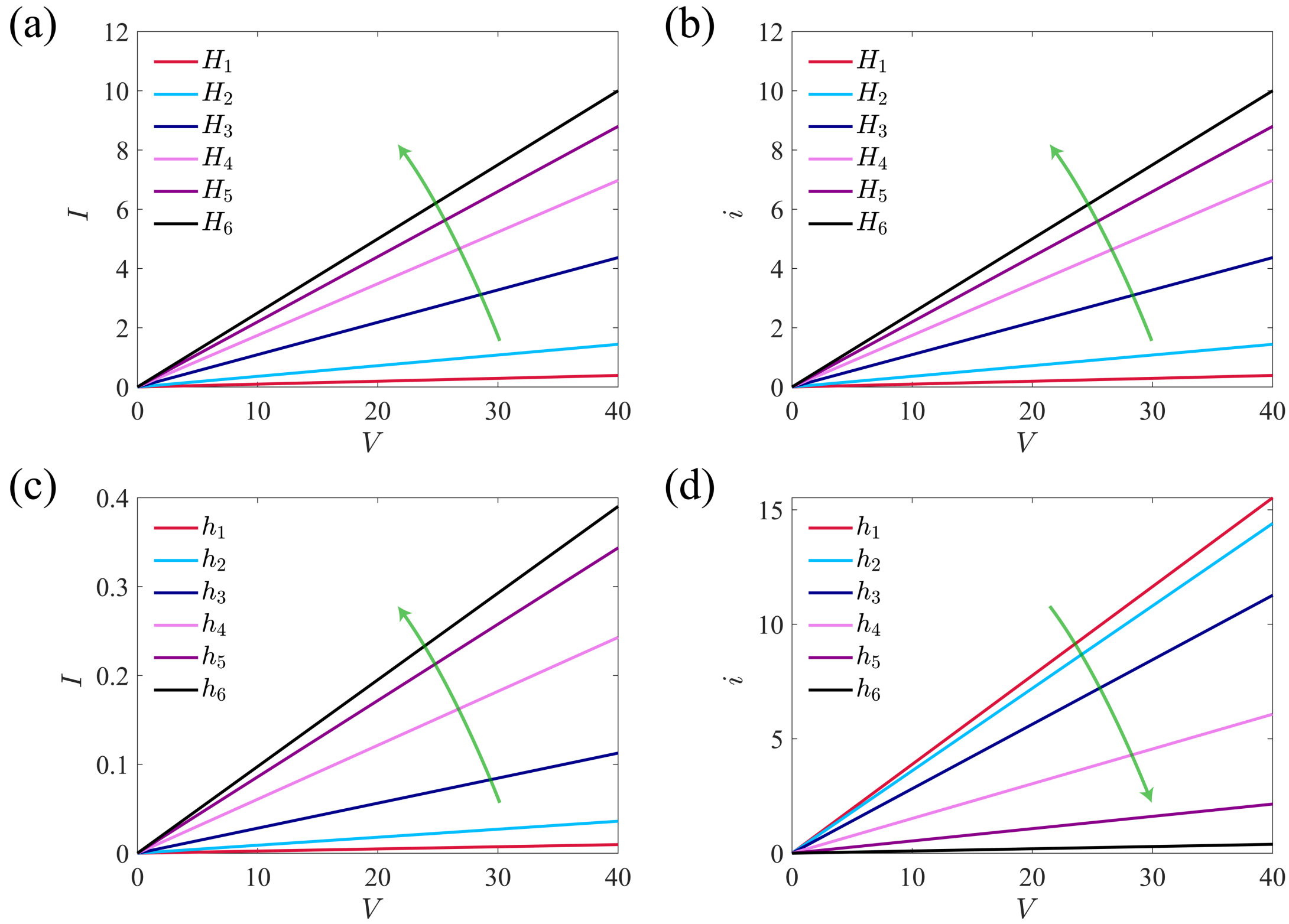

**Figure 9** Variation of electric response for a unit-cell operating at the limit of vanishing selectivity [Eq. (6)] for a square microchannel ($H = W$) and square nanochannel ($h = w$): (a) current-voltage, $I - V$, and (b) current-density-voltage response, $i - V$ for a unit-cell where the nanochannel geometry is kept constant and the microchannel geometry is varied. Similarly, the (c) $I - V$ and (d) $i - V$ for constant microchannel geometry and varying nanochannel geometry. In all figures, the green arrow points in the direction of increasing size. Note that currents, $I$, current densities, $i$, voltages, $V$, and geometric parameters, $H$ and $h$ are dimensionless. The default geometry of the unit-cell, given by subscripts '1', is given in **Table S1** of the Supporting Information[22].

Wang et al.[43] attribute the changes to concentration polarization that appear upon upscaling from a single channel to a multichannel system. We emphasize that the effects of concentration polarization are always apparent – regardless of the number of channels in the system. Note that Eq. (2) [as well as Eqs. (6) and (7)] are derived from the PNP equations where the effects of concentration polarization are inherently manifested. In fact, all resistances are due to concentration polarization.

To better understand the issue of currents versus current densities, we can consider either Eq. (7), which is relevant to RED processes, or Eq. (6), which is not relevant to RED but has the added benefit that it is surface charge independent. Regardless of the scenario, to increase the ionic currents, one must decrease the resistance as much as possible. Hence, one should decrease $\Sigma R_{\text{micro}}$ as much as possible. This suggests reducing $R_{\text{FF}}$ as much as possible. In the limiting case, one has $R_{\text{FF}} = 0$, when the ratios $w/W$ and $h/H$ approach unity. This scenario corresponds to a 1D membrane system –suggesting that membranes are the ultimate tool for large-scale RED. Here, the ratio $S_{\text{nano}}/S_{\text{micro}}$ is maximal, allowing for enhanced fluxes. However, one still needs to account for the effects of $R_{\text{micro}}$ which is often neglected. From the practical standpoint, in multichannel systems, one can never approach $S_{\text{nano}}/S_{\text{micro}} \rightarrow 1$.Yet, Siria et al.[5] are correct in stating that isolated systems have larger currents and current densities. We now demonstrate this.

In **Figure 9**, we consider a cuboidal microchannel and cuboidal nanochannels where $w = h$. **Figure 9**(a) considers a scenario where the nanochannel geometry is kept constant (i.e., $w = h =$ *const.*) and the microchannel geometry is varied. Observe that as $H$ grows, the nanochannel becomes more isolated from its neighbors, and the current increases. This is because $R_{\text{micro}}$ is decreasing while $R_{\text{FF}}$ is reaching its lowest value (the square equivalent of $R_{\text{access}}$). Naturally, dividing the current by the constant nanochannel area does not change the trend of the current density, $i = I/S_{\text{nano}}$ [**Figure 9**(b)]. Physically, increasing the microchannel geometry flanking a nanochannel of a given geometry leads to higher ionic fluxes. **Figure 9**(c) considers a scenario where the microchannel geometry is kept constant (i.e., $W = H = const.$) and the nanochannel geometry is varied. In contrast to the previous scenario, in **Figure 9**(d) it can be observed that as $S_{\text{nano}}$ is increased, the current density decreases such that the highly isolated channels have larger current densities and closely packed channels have smaller current densities. The findings related to optimizing the current and current densities can be summarized rather succinctly:

- Enhanced current densities occur for highly isolated systems [This corresponds to a case where **Figure 1**(e) has a large area of grey membrane material]. However, the harvested current is limited by the total size of the system.
- Large total currents appear in highly packed (and non-isolated) systems. In such systems, while the current is maximized, the current density is correspondingly minimized.

Hence from a practical point of view, one should define a priori what one is trying to maximize, given the additional constraints of space, thermal and mechanical stability.

### 4.4. Accounting for Surface Charge Regulation

In recent years there has been an increased interest in the effects of surface charge regulation (SCR), whereby the surface charge (density) can be modulated by adsorption dynamics of ions onto the surface[38,39,41,44–47]. Briefly, in these works, it is shown that the surface charge density is a function of the bulk concentration, $c_0$, pH, and other system parameters, such that $\sigma_s(c_0) \sim c_0^{\alpha}$. Since the average excess concentration, $\Sigma_s$, is linear with the surface charge density [Eq. (5)], the nanochannel resistance [$R_{\text{nano}} c_0/\Sigma_s$ in Eq. (7)] is modified accordingly. Since Eq. (2) is the exact solution to the PNP equations, it is relatively straightforward to incorporate SCR models[41,48,46] into the solution for $R_{\text{unit-cell}}$ by modifying $\Sigma_s$. This is another added advantage of our theory.

**Figure 10** demonstrates how the total resistance changes when microchannels (full unit-cell) and SCR are accounted for. First, we plot the nanochannel-only resistance, $R_{\text{unit-cell}}^{(\Sigma R_{\text{micro}}=0)}$ [Eq. (10)] without SCR – this is the solid red line with a slope of $\alpha = 0$ at low concentrations. The effect of SCR is that the slope at low concentration is now $-\alpha$ (blue dashed line). The dotted magenta line denotes $R_{\text{unit-cell}}$ [Eq. (2)] which accounts for $\Sigma R_{\text{micro}}$. Note, again, that at low concentrations, the resistance no longer saturates to a constant value, but rather the slope is -1 due to $\Sigma R_{\text{micro}}$. When SCR is added to Eq. (2) we have a more complicated result – this is the dashed-dotted dark-blue line. At high concentrations, the response is determined by the bulk nanochannel resistance with a slope of -1. At intermediate values, the response is still determined by the nanochannel, but now this resistance is determined by SCR such that the slope is $-\alpha$. At even lower concentrations, $\Sigma R_{\text{micro}}$ dominates the response and the curve achieves a slope of -1.

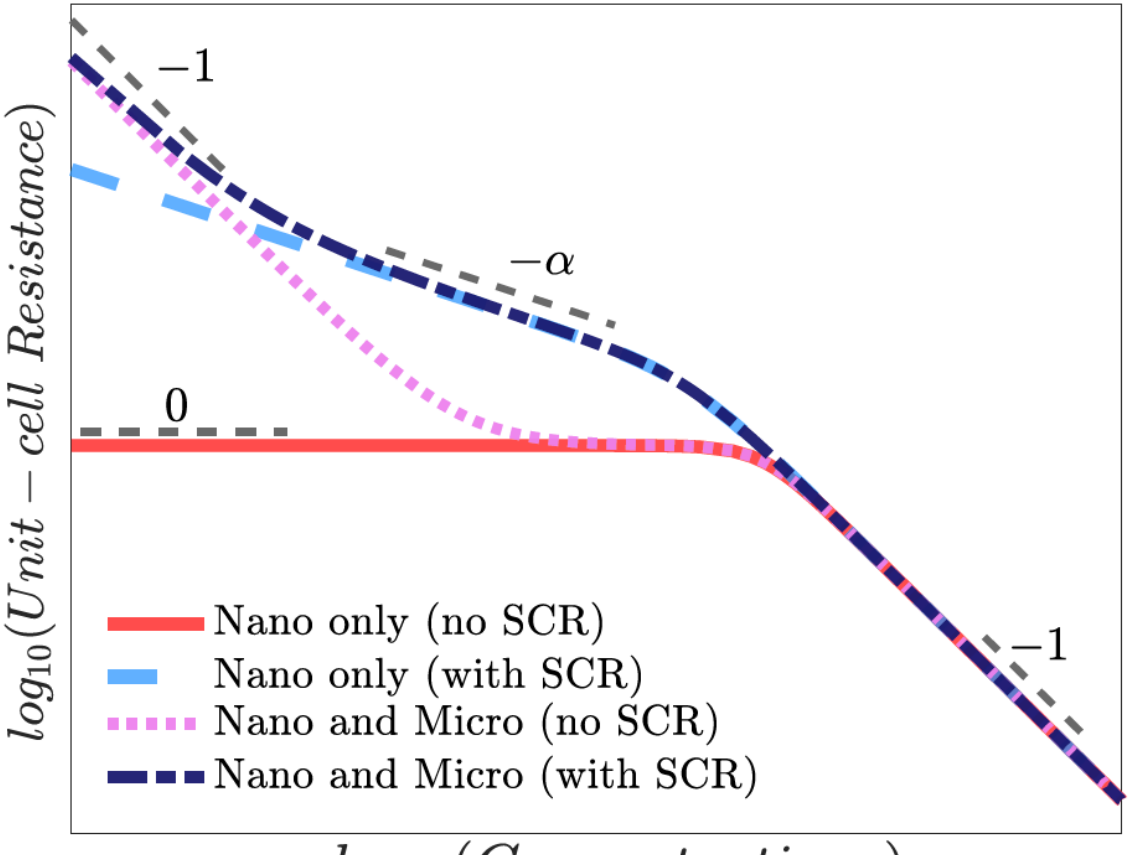

**Figure 10** A schematic of the unit-cell resistance versus the concentration without and with the effects of the microchannels and surface-charge regulation. The nanochannel-only resistance, $R_{\text{unit-cell}}^{(\Sigma R_{\text{micro}}=0)}$, is given by Eq. (10) , while the unit-cell response, which accounts for the microchannels, is given by Eq. (2). The effects of surface charge regulation are added, as discussed in the main text.

### 4.5. Beyond the ohmic response

#### *4.5.1. Concentration polarization*

In this work, we have focused solely on the ohmic response where the current and voltage are linearly related. We have done this because it is easier and more intuitive but also because it is a universal response that holds for all values of $\Sigma_s/c_0$. In particular, when $\Sigma_s/c_0 \ll 1$, channels are vanishingly selective such that $V = IR_{\text{unit-cell}}^{(\text{vanishing})}$ [Eq. (6)] for all values of the current or the voltage. However, when $\Sigma_s/c_0 \gg 1$, the response is by far more complicated. At low currents (and low voltages), the response is linear, as given by Eq. (7), $V = IR_{\text{unit-cell}}^{(\text{ideal})}$. However, at higher voltages the current reaches a limiting (diffusion-limited) value. This is because the voltage has a logarithmic dependence on the current.[30,49] The appearance of the limiting current is often attributed to the appearance of concentration polarization – however, it should be emphasized that Eqs. (2)-(7) can be derived from either the general PNP equations [Eq. (2)] or reduced PNP equations (these are the PNP equations solved under the limiting assumption of highly [Eq. (7)] or vanishingly [Eq. (6)] selective systems). In other words, regardless of $\Sigma_s/c_0$, and regardless of the voltage (low or high), concentration polarization is always manifested in these systems. However, only in certain regimes do the effects of concentration polarization become strong enough such that depletion and enrichment layers become apparent, resulting in limiting currents. In our past work[23], we provide the general $I-V$ response for arbitrary values of $\Sigma_s/c_0$. However, there $I$ also depends on the salt current density, $J$, such that $I(J)$ needs to be evaluated numerically from a transcendental equation. We note that another future direction will be to understand whether the unit-cell concept holds, and if so, how, in the limiting current regime. Finally, we note that limiting currents constitute a relevant regime in the operation of fuel cells[50] – however, a comprehensive discussion on the similarities and differences between fuel cells and nanofluidics is beyond the current scope of this work. In the following sub-section, we will discuss how the limiting current can be surpassed through the effects of electroconvection.

#### *4.5.2. Effects of electroconvective transport*

Electroconvection can manifest itself in nanofluidics systems in two manners. First, there can be an additional directed advective conductance that increases the total electrical conductance of the nanopore/nanochannel by advection of ions across the membrane or nanochannel. This increase can be rather substantial; however, here we have neglected it. The reason for this is two-fold. First, the realistic nanoporous membrane is tortuous and includes many channel intersections and dead ends. Often, this leads to the decay of the advective current. Second, in contrast, to a single pore system (with a simple geometry), the advective contribution can be calculated only within the nanochannel (when the effects of the microchannels are neglected). The advective conductance within a three-layered system (i.e., microchannel-nanochannel-microchannel) is an open question.

The second manifestation occurs within the microchannel at the nanochannel-microchannel interface, whereby an electroconvective instability appears at the interface. Briefly, above a critical voltage (after the limiting current has already appeared), the interfacial space charge layer loses stability such that a non-linear electric body force drives the fluid motion. Typically, this instability is observed when a large number of small vortices form at the interface. Over time, the size of the vortices increases while the number decreases. Eventually, an almost steady array of vortices is formed. Importantly, this is a well-established mechanism[10,40,51–54] that has been shown to surpass the diffusion-limited limiting current regime – this is the overlimiting current regime[23,49]. Very little work has been conducted on researching how this instability is manifested in an array of nanochannels (which is essentially different than the macroscopically large array).

One thing is clear regarding the effects of electroconvection in nanochannel arrays – when instability appears – the concept of the unit-cell (i.e., no exchange of flux across symmetry planes) breaks down such that most of our results no longer hold. While it might appear that our results are limited to the Ohmic regime, it should be pointed out that the general trends of the results predicted by this work have been shown to experimentally hold for limiting currents as well as over-limiting currents[33]. Thus, while we are unable to explain all the exact details of a realistic system, under the effects of the electroconvective instability, we are indeed able to shed light on the underlying first principles. A coherent theory for the advective transport of ions in multiple channel systems remains an elusive open question.

## 5. Conclusions

This work addresses the open question of relating the electrical response of a single nanochannel system to the response of an array. Understanding and elucidating this upscaling is of particular importance to the applicability of nanochannel arrays used for desalination and energy harvesting systems. Here, we propose to treat the random nanoporous membrane geometry as a simplified array whose analysis can be conducted straightforwardly. Notably, the results provide remarkable physical insights – namely, that nanochannel arrays can be represented as a simple parallel electrical circuit comprised of a unit-cell resistance.

Our starting point is the microchannel-nanochannel array presented in **Figure 2**(a), whereby we observe that it is comprised of "unit-cells" [**Figure 2**(b)]. The resistance of each unit-cell, $R_{\text{unit-cell}}$ [Eq. (2)] can then be subdivided into three separate contributions [**Figure 2**(d)]: 1) nanochannel resistance, $R_{\text{nano}}$, 2) microchannel resistance, $R_{\text{micro}}$, and 3) field focusing resistance, $R_{\text{FF}}$ (a generalization of access resistance, $R_{\text{access}}$). We provide a new analytical expression for $R_{\text{FF}}$ which holds for nanochannels of arbitrary cross-section in arbitrary confinement. We then show that the total resistance, $R_{\text{total}}$, is that of an electrical circuit comprised of $N$ unit-cell connected in parallel such that $R_{\text{total}} = R_{\text{unit-cell}}/N$ [**Figure 2**(d), Eq. (1)]. While this appears to be relatively intuitive and utilizes the simplest of electrical circuits imaginable, to our dismay and surprise, to the best of our knowledge, the model presented in this work [and confirmed by numerical simulations (Secs 3.1) and experiments (Sec. 3.2)] has not been previously presented or discussed.

A few important comments are warranted. First, while it may appear that we have used a simple abstraction to represent the electrical circuit – this is not the case. In fact, $R_{\text{unit-cell}}$ [Eq. (2)] represents the exact solution to the Poisson-Nernst-Planck equations in the appropriate limit of Ohmic conductance. Hence, the abstraction presented in this work is not merely another hypothesized electrical circuit – it is the accurate equivalent circuit, and it supersedes all other suggested models (Sec. 4). Second, another consequence of this model is that the analysis of a multichannel system is now reduced to the problem of studying the much simpler unit-cell problem, when one or more types of unit-cells are connected in parallel. Third, this work has focused on the Ohmic resistance (or Ohmic conductance); however, so long as the effects of electroconvection remain suppressed, the current-voltage response can be generalized to larger voltages when the currents are diffusion-limited[52,55]. Fourth, without going into additional details given in the above discussion (Sec. 4), this model provides theoretical predictions that can rationalize previous experimental works.

Thus, our approach of electrical circuit modeling and the underlying theoretical models serve as a tractable analytical method to aid the accurate interpretation of experiments, analysis of ED/RED systems of scale, and the design of specific nanofluidic circuitry.

### Acknowledgments

This work was supported by the Israel Science Foundation (Grant Nos. 337/20 and 1953/20). We thank the Ilse Katz Institute for Nanoscale Science & Technology for their support.